\documentclass[12pt]{iopart}
\usepackage{iopams}
\usepackage{epsf}
\begin{document}

\letter{IV-VI resonant cavity enhanced photodetectors for the
midinfrared}

\author{M. B\"oberl, T. Fromherz, T. Schwarzl, G. Springholz and W.
Heiss \footnote[1]{corresponding author:
\mailto{wolfgang.heiss@jku.at}} }

\address{Institut f\"ur Halbleiter- und Festk\"orperphysik,
Universit\"at Linz, Altenbergerstr. 69, 4040 Linz, Austria}

\begin{abstract}
A resonant-cavity enhanced detector operating in the mid-infrared
at a wavelength around 3.6 $\mu$m is demonstrated. The device is
based on a narrow-gap lead salt heterostructure epitaxially grown
on a BaF$_2$(111) substrate. Below 140 K, the photovoltage clearly
shows a single narrow cavity resonance, with a
$\Delta\lambda/\lambda$ ratio of only 2 \% at 80 K.

\end{abstract}



\maketitle

\section{Introduction}
The sensitivity for detection of trace gas molecules in air or in
other carrier gases by measuring the characteristic fingerprint
like vibration-rotation absorption lines of molecules is in the
spectral range of the mid-infrared by far higher than in the near
infrared or in the visible. Thus, for sensitive gas analysis and
atmospheric pollution monitoring highly efficient optoelectronic
devices for the mid-infrared are required. Narrow-gap IV-VI
semiconductors are promising candidates as mid-infrared detectors;
polycrystalline IV-VI photodetectors operating at room temperature
are even commercially available. In addition, two-dimensional PbTe
focal plane arrays on a Si-substrate were realized recently
\cite{alchalabi01}. For molecular spectroscopy, the emitter or
detector or both have to be tuned to the molecule absorption line.
This requisite can be achieved by either using a narrow band
emitter and a broadband detector or a broadband source such as a
glowbar and a detector with a small spectral bandwidth. In
contrast to the above mentioned lead salt photodetectors, which
have a relatively large spectral bandwidth, we demonstrate a
photodetector with a small spectral bandwidth. In this so-called
resonant cavity enhanced photodetector (RCEPD) a thin absorbing
layer is placed in a vertical optical cavity. This yields
advantages compared to conventional detectors because (a) the
narrow detection wavelength of interest can be tuned by the cavity
length and (b) the quantum efficiency is enhanced due to the
standing wave effect caused by the cavity\cite{unlu90}. Up to now
most work has been founded on RCEPDs operating in the near
infrared for telecommunication purposes. These structures were
fabricated from SiGe/Si layers \cite{cheng02} and from GaAs-based
heterostructures \cite{heroux99,kinsey00}. In the midinfrared
first attempts on RCEPDs based on CdHgTe compounds
\cite{pautrat97} or fabricated of InAs \cite{green03} were demonstrated. These devices operate around 3.1 $\mu$m.\\
In this work, we demonstrate a RCEPD based on the narrow gap lead
salt semiconductors operating at a wavelength of $\lambda$ = 3.6
$\mu$m.

\section{Experiments}

\begin{figure}
\begin{center}
\epsfbox{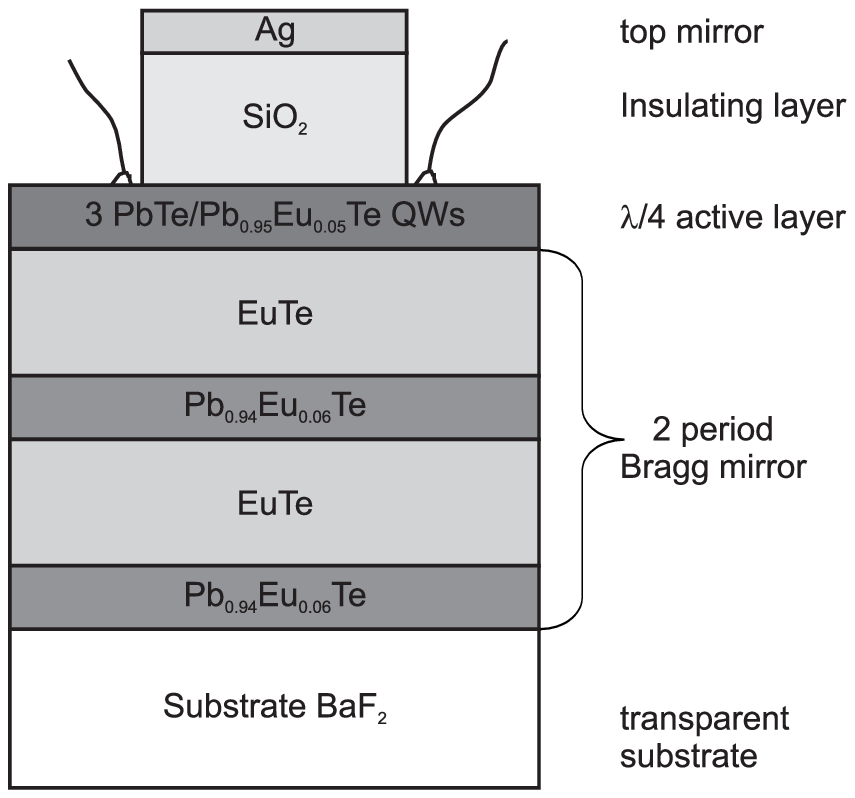}
\end{center} \caption{\label{structure}Schematic
sketch of the resonant-cavity photodetector structure.}
\end{figure}

\begin{figure}
\begin{center}
\epsfbox{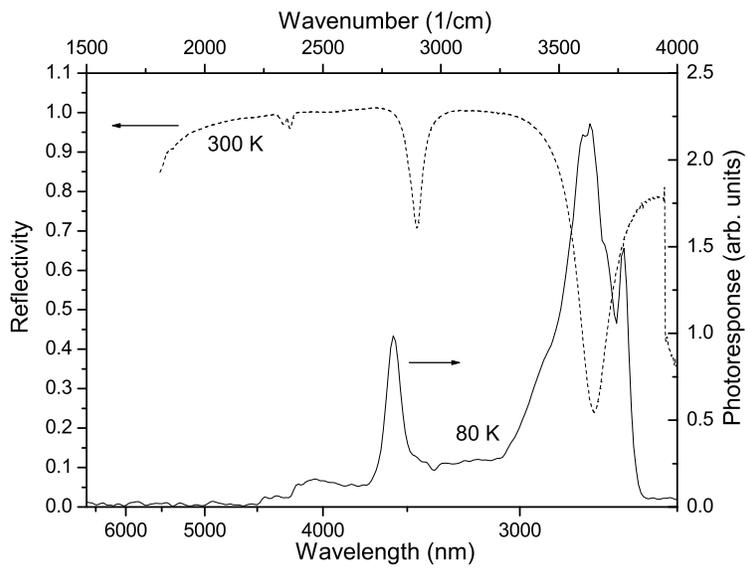}
\end{center}\caption{\label{spectra}Room
temperature reflectivity
 spectrum of the RCEPD structure (dashed line). Photovoltage of the structure
measured at 80 K (solid line).}
\end{figure}

Figure \ref{structure} shows the schematic structure of the
resonant cavity photodetector. The multilayer detector sample is
grown by molecular beam epitaxy on (111)-orientated BaF$_2$ that
is transparent in the mid-infrared. The design of the lead salt
RCEPD is similar to the design of IV-VI vertical-cavity
surface-emitting laser structures \cite{furst02,furst04}. However,
the RCEPD structure consists of only one Bragg mirror with a
metallic mirror as second mirror. In detail, a two-period Bragg
interference mirror consisting of $\lambda$/4 thick
Pb$_{0.94}$Eu$_{0.06}$Te/EuTe layer pairs acts as bottom mirror of
the sample. The Pb$_{0.94}$Eu$_{0.06}$Te mirror layers are grown
as digital alloy of PbTe and Pb$_{0.59}$Eu$_{0.41}$Te with a
superlattice period of 52 \AA. Due to the high refractive index
contrast of more than 80 \% \cite{heiss01} between the two mirror
materials, the two-period Bragg mirror exhibits a reflectivity as
high as 98 \%. The active region is composed of 3
PbTe/Pb$_{0.95}$Eu$_{0.05}$Te quantum wells (QWs) with a PbTe well
width of 50 \AA\ and a Pb$_{0.95}$Eu$_{0.05}$Te barrier width of
300 \AA. On top of the absorbing region, a 1.2 $\mu$m thick
SiO$_2$ layer serving as insulator is deposited using
plasma-enhanced chemical vapor deposition. The total length of the
cavity complies with 3$\lambda$/4, so that only one single
resonance appears in the Bragg mirror stopband region. Mesas are
fabricated using standard photolithography and wet chemical
etching. On top of the mesas an Ag layer is sputtered serving as
top cavity mirror. The active layer is contacted close to the top
mirror mesa ensuring the optimal use of the resonant cavity
effect. Gold wires are bonded on In pads forming a Schottky
contact. For photoresponse experiments, the sample was mounted in
a He flow cryostat. The samples were illuminated under normal
incidence from the transparent substrate side using a glowbar.
After passing a chopper and an optical low pass filter with a
cut-off wavelength of 2.63 $\mu$m, the radiation was focused on
the sample. Photovoltage measurements were performed with a
Fourier-transform infrared (FTIR) interferometer, which was
operated in step-scan mode measuring the photovoltage by a
standard lock-in technique. For all experiments, the device was
operated under open-circuit conditions.
\\\\
The structure was first characterized by FTIR reflectance
experiments. The reflectivity spectrum of the sample measured at
room temperature is shown as dashed line in Figure \ref{spectra}.
It exhibits one clear cavity resonance of m=2$^{nd}$ order due to
the cavity length of three-quarter of $\lambda$. The resonance
peak is at 3.45 $\mu$m and has a full width at half maximum (FWHM)
of about 70 nm which results in an effective finesse of 24. As
shown in Figure \ref{spectra}, the width of the broad Bragg mirror
stopband is more than 2 $\mu$m owing to the high refractive index
contrast of the mirror materials \cite{heiss01}. At the
low-wavelength edge of the stopband (2.47 $\mu$m) an interference
fringe from the total thickness of the structure appears. The two
small dips in the reflectivity at 4.2 $\mu$m are absorption lines
from the remaining CO$_2$ in the N$_2$-purged sample chamber.\\

In Figure \ref{spectra}, the photoresponse at 80 K of the device
operating in photovoltaic mode is shown. The signal peak at 3.57
$\mu$m corresponds to the position of the cavity resonance at 80
K. It is slightly shifted to higher wavelengths with respect to
the cavity resonance observed in the reflectivity spectrum at room
temperature due to the increase of the refractive index of the
cavity materials with decreasing temperature. The linewidth of the
cavity resonance in the photoresponse signal is only 80 nm
corresponding to a small wavelength ratio $\Delta\lambda/\lambda$
of 2 \%. This linewidth is slightly larger than the width of the
cavity resonance in the reflectance spectrum. This is due the fact
that at 80 K, the active layer (the three thin PbTe QWs) acts as
an absorber at the wavelength of the cavity resonance, while this
is not the case at room temperature due to the strong temperature
dependence of the PbTe bandgap. Therefore, at 80 K the cavity
resonance is slightly damped and broadened leading to an increased
linewidth in the
photovoltage spectrum.\\
A small photovoltage signal is  detected in the stopband region
around the resonance (between the onset of the QW absorption at
4.5 $\mu$m and 3 $\mu$m). This signal should be suppressed by the
high reflectivity of the bottom Bragg mirror. However, because the
reflectivity of the Bragg mirror is only 98 \% not all of the
incoming light is reflected back. At smaller wavelengths than the
stopband region ($\lambda$ smaller than 3.0 $\mu$m) also a strong
photovoltage signal is found. We assume that the
Pb$_{0.94}$Eu$_{0.06}$Te layers of the Bragg mirror which absorb
in this wavelength region contribute to the photovoltage signal.
The cut-off of the photovoltage at 2.63 $\mu$m is due to the used
low-pass filter in
the experimental setup.\\
Detector operation is observed up to 140 K. For the detector
design we used in this work, the resonance in the photoresponse
signal is quenched at 140 K, because the bandgap of the
PbTe/Pb$_{0.95}$Eu$_{0.05}$Te QWs shifts to higher energies than
the energy of the cavity resonance. This is due to the fact that
the bandgap of the active material increases much stronger with
increasing temperature than the refractive index decreases
\cite{springholz00}. Designing the detector structure for a cavity
resonance at higher energies, should allow obtaining detector
operation temperatures reached by simple thermoelectric coolers.

\section{Conclusions}
We showed a lead salt resonant-cavity enhanced photodetector based
on a two-period Bragg mirror and a metallic mirror and 3
PbTe/PbEuTe QWs as absorbing material in between. The operation
wavelength at 80 K is 3.57 $\mu$m and the $\Delta\lambda/\lambda$
ratio is only 2 \%. By changing the cavity length, the resonance
position can be tuned to a certain molecular absorption line,
making such photodetectors a promising device for molecular gas
sensing applications. This demonstration of the
resonant-cavity-effect from a non-optimised structure gives
prospect to gain suitable lead-salt photodetectors for gas
absorption applications in the near future.

\ack The authors are grateful to S. Andreeva for technical
assistance. Financial support from FWF (projects Y179 and P15583)
and GME of Austria is gratefully acknowledged.

\Figures

\end{document}